\documentstyle[epsf]{elsart}

\begin{document}

\hfill TK 95 31

\bigskip 

\begin{frontmatter}
\title{CHIRAL LAGRANGIANS FOR BARYONS COUPLED
  TO MASSIVE SPIN--1 FIELDS \thanksref{dfg}}
\thanks[dfg]{Work supported in part by the Deutsche Forschungsgemeinschaft}


\author[Bonn]{B.~Borasoy}\footnote{email: borasoy@pythia.itkp.uni-bonn.de},
\author[Bonn]{Ulf-G.~Mei\ss ner}\footnote{email:
                                          meissner@pythia.itkp.uni-bonn.de}


\address[Bonn]{Institut f\"ur Theoretische Kernphysik, Universit\"at
  Bonn\\ Nu\ss allee 14--16, D-53115 Bonn,  Germany}


\begin{abstract}
We analyze the effective low--energy field theory of Goldstone bosons
and  baryons chirally coupled to massive spin--1 fields. We use
the electromagnetic baryon form factors to demonstrate the formal
equivalence between the vector and the tensor field formulation for
the spin--1 fields. We also discuss the origin of the so--called
Weinberg term in pion--nucleon scattering and the role of
$\rho$--meson exchange. Chirally coupled vector mesons do not give
rise to this two--pion nucleon seagull interaction but rather to
higher order corrections.  Some problems of the formal equivalence
arising in higher orders and related to loops are touched upon.
\end{abstract}

\end{frontmatter}


\section{Introduction}
Over the last years, there has been considerable activity in the
formulation of effective chiral Lagrangians including baryons, see
e.g. \cite{gss,jm,dobo,bkkm} or the recent review \cite{bkmrev}. Beyond
leading order, such effective Lagrangians contain coefficients not
fixed by symmetry, the so--called low--energy constants (LECs)
\cite{wein79,gl84}. In general, these  LECs should  be determined from
phenomenology. However, in the baryon sector, there do not yet exist
sufficiently many accurate data which allow to independently determine
all these LECs. That is quite different from the meson sector. There,
Gasser and Leutwyler \cite{gl84,gl85} performed a systematic study of
S-matrix  and transition current matrix elements to pin down
the LECs. It was already argued in Ref.\cite{gl84} that the numerical
values of certain LECs can be understood to a high degree of accuracy
from vector meson exchange. This picture was later sharpened by Ecker
et al. \cite{reso,reso1} who showed that {\it all} of the mesonic LECs are
indeed saturated by resonance exchange of vector, axial--vector,
scalar and pseudoscalar type (some of these aspects were also
discussed in \cite{reso2}). However, there is a potential problem
related to the massive spin--1 fields. In Ref.\cite{reso}, an
antisymmetric tensor field formulation well--known in supergravity was
used to represent these degrees of freedom. One could, however, choose
as well a more conventional vector field approach (for a review, see
\cite{ulfpr}).  At first sight, these two formulations lead to
very different results, like e.g. the pion radius is
zero and non-vanishing for the vector and tensor field formulation,
respectively. In Ref.\cite{reso1}, high energy constraints were used
to show that there are local terms at next--to--leading order in the
vector field approach which lead to a full equivalence between the two
schemes, as intuitively expected. A path integral approach, which
gives a formal transformation between the two representations, has also
been given recently \cite{bp} (we give some clarification in appendix B).

In the baryon sector, the situation is in a  much less satisfactory
state. Only recently, the renormalization of all divergences at order
$q^3$ (where $q$ denotes a small momentum or meson mass) has been
performed \cite{eckerr} and all finite terms at that order have been
enumerated \cite{eckmoj} (in the heavy baryon formalism and for two
flavors). Previously,
Krause \cite{krause} had listed all terms up--to--and--including
${\cal O}(q^3)$ in three-flavor relativistic baryon CHPT.
The best studied baryon chiral perturbation
theory LECs are the ones related to the dimension two Goldstone boson
($\phi$)--baryon ($B$) Lagrangian ${\cal L}_{\phi B}^{(2)}$ for the
two flavor case, see e.g. \cite{bkmrev,bkmpin,bkmppn}. These can be
understood from resonance exchange saturation as detailed in
\cite{ulfmit}, but not to the same degree of accuracy as it is the
case in the meson sector. The concept of resonance exchange saturation
of the low--energy constants is more complicated when baryons are
present. One does not only have to consider $s$--channel baryon
excitations (like the $\Delta (1232)$,\footnote{We are not considering
the decuplet states as dynamical degrees of freedom in the effective
field theory here.} the $N^* (1440)$, and so on) but also $t$--channel
meson excitations (like scalars, vectors or axials). Therefore, the
problem of how to couple the spin--1 fields to the baryons
arises. This was already mentioned by Gasser et al. \cite{gss} and
further elaborated on in Ref.\cite{ulfsea}. In the framework of
relativistic baryon CHPT, one finds a LEC contributing to the
isovector magnetic moment very close to zero, at odds with the
expectation of a large $\rho$--meson induced higher dimension
operator. It is thus appropriate to address the problem of the
equivalence of the vector and tensor field formulation for the spin--1
mesons in the presence of baryons (which are treated relativistically
here. Only after calculating the spin--1 field contribution to a
certain LEC, one is allowed to take the heavy fermion limit for the baryons).
Here, we will show how to couple massive spin--1 mesons to the
effective Goldstone boson--baryon Lagrangian consistent with chiral
symmetry and demonstrate that physics does not depend on the
particular choice of representation for the spin--1 fields.

There is, however, yet another motivation. In many meson--exchange
models of the nulcear force, one includes vector meson exchange. The
$\rho$--contribution is believed to suppress the pion tensor force at
small and intermediate separations whereas the $\omega$ makes up part
of the short range repulsion. If chiral symmetry is imposed on such
models, one also has a set of local operators involving pions and
nucleons, in particular the two--pion--nucleon seagull, the so--called
Weinberg term, with a coupling strength $\sim 1/F_\pi^2$ (with $F_\pi =
93$ MeV the pion decay constant). On the other hand, it is well known
that treating the $\rho$--meson as a massive Yang--Mills boson,
$\rho$--exchange in fact produces the Weinberg term with the right
strength, see Fig.~1. There appears to be a double--counting
problem. Let us therefore elaborate a bit on this. First, the chiral
covariant derivative term acting on the nucleon fields $\Psi$ (for the
moment, we restrict ourselves to the two--flavor case of pions and nucleons),
\begin{figure}[t]
\hskip 1.2in
\epsfysize=1.5in
\epsffile{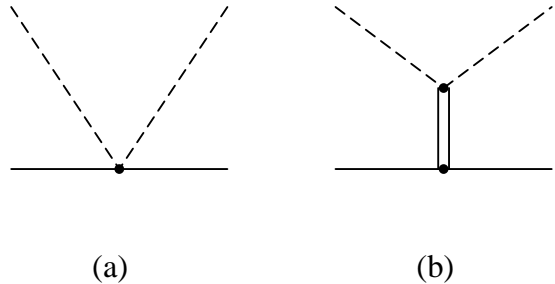}
\caption{\label{fig1} The two--pion nucleon seagull (Weinberg
term). (a) shows the fundamental Feynman diagram (from the chiral
covariant derivative) and  (b) represents its generation via
$t$--channel $\rho$--meson exchange. Solid and dashed
lines denote the octet baryons and Goldstones, respectively.}
\end{figure}
\begin{equation}
{\cal L}_{\phi B} = i \, \bar{\Psi} \, \gamma_\mu D^\mu \, \Psi + \ldots =
i \, \bar{\Psi} \, \gamma_\mu \, (\partial^\mu + \Gamma^\mu ) \, \Psi + \ldots
\label{chicov}
\end{equation}
with $\Gamma_\mu$ the chiral connection, contains besides many others,
the Weinberg term with the Feynman insertion (i.e. the nucleon and pion
fields are not shown)
\begin{equation}
\frac{1}{4 F_\pi^2} \gamma_\mu (q_1 + q_2)^\mu \, \epsilon^{abc}
\, \tau^c
\label{weinterm}
\end{equation}
with $q_i$ the pion momenta and '$a,b,c$' are isospin indices. This
term leads to a low--energy theorem (LET) for the isovector
pion--nucleon scattering  amplitude $T^-$ in forward direction,
\begin{equation}
\frac{T^-(\nu )}{ \nu } \to \frac{1}{F_\pi^2} \, \, \, ,
\label{let}
\end{equation}
as $\nu \to 0$ and $\nu = p \cdot q \, / \, m$ the standard momentum transfer
variable ($p$ denotes the nucleon momentum).
On the other hand, conventional $\rho$--meson exchange with
the following insertions
\begin{equation}
\rho NN: \, \, \,  -g_{\rho NN} \, \epsilon \cdot \gamma \,\frac{\tau^c}{2}
\, \, , \quad
\rho \pi \pi: \, \, \, g_{\rho \pi \pi} \, i \, \epsilon^{bac} \, \epsilon^\nu
\, (q_1 + q_2)_\nu \, \, \,  ,
\label{rhocoup}
\end{equation}
where $\epsilon_\nu$ denotes the polarization vector of the spin--1
field, leads to
\begin{equation}
\lim_{\nu \to 0} \frac{T^-(\nu )}{ \nu } = 2 \, g_{\rho NN} \,
g_{\rho \pi \pi} \frac{1}{M_\rho^2} = \frac{1}{F_\pi^2} \, \, \, ,
\label{rhowein}
\end{equation}
using universality ($g_{\rho NN} = g_{\rho \pi \pi} \equiv g$) and the KSFR
relation \cite{kfsr}, $M_\rho^2 = 2 g^2 F_\pi^2$, which is fulfilled
within a few percent in nature. So now there seem to be two sources to give
the low--energy theorem, Eq.(\ref{let}).\footnote{We are grateful to
Harry Lee for reminding us of this problem. Of course, it has already
been solved long time ago but using another approach than the one
advocated here, see e.g. Weinberg' s paper \cite{weinnon} and also
section 4.}  In what follows, we will show
that constructing the couplings of the spin--1 fields to Goldstone
bosons and baryons solely based on the non--linear
realization of chiral symmetry, $\rho$--exchange does not contribute to
the LET but only to some of its higher order corrections. The Weinberg
term will thus entirely arise from the nucleon kinetic term,
Eq.(\ref{chicov}), i.e. from the chiral covariant derivative acting on
the nucleon fields (see also the footnote in ref.\cite{bkmpin}).

The paper is organized as follows. In section 2, we briefly summarize
the effective Goldstone boson--baryon Lagrangian. We discuss in some
detail the couplings of massive spin--1 fields in the vector and the
tensor field formulation. We also discuss some formal equivalence
between these two schemes making use of the Proca equation. In section
3, we calculate the electromagnetic form factors of the ground state
baryon octet. This allows to fix the coefficients of the contact terms
one has to add to the vector formulation so as to achieve complete
equivalence to the tensor one up--to--and--including order $q^3$. In
section 4, we return to the question of $\rho$--exchange in
pion--nucleon scattering and show that chirally coupled vector mesons
do not interfere with the leading order Weinberg term. Some open
problems related to higher orders, in particular the effects of pion
loops,  are addressed in section 5. A short
summary is given in section 6. Some technicalities related to the
construction of the effective Lagrangian are relegated to appendix A.
Appendix B contains some comments on the formal equivalence proof
based on dual transformations.

\vfill

\section{Effective Lagrangian of Goldstones, Baryons and Spin--1 Fields}

To perform the calculations, we make use of the effective
Goldstone boson--baryon Lagrangian. Our notation is identical to the one
used in \cite{bkmz} and we compile here only some necessary definitions.
Denoting by $\phi$ the pseudoscalar Goldstone fields ($\pi, K, \eta$) and
by $B$ the baryon octet ($N,\Lambda,\Sigma,\Xi$),
the  effective Lagrangian takes the form
(to one loop accuracy)
\begin{equation}
{\cal L}_{\rm eff} = {\cal L}_{\phi B}^{(1)} +  {\cal L}_{\phi B}^{(2)} +
  {\cal L}_{\phi B}^{(3)} +
  {\cal L}_{\phi B}^{(4)} + {\cal L}_{\phi}^{(2)}+ {\cal L}_{\phi}^{(4)}
\label{leff}
\end{equation}
where the chiral dimension $(i)$ counts the number of derivatives
and/or meson mass insertions. At this stage, the baryons are treated
relativistically. Only after the resonance saturation of the LECs
has been performed, one is allowed to go to the extreme
non--relativistic limit \cite{jm,bkkm}, in which the baryons are characterized
by a four--velocity $v_\mu$. In that case, there is a one--to--one
correspondence between the expansion in small momenta and quark masses
and the expansion in Goldstone boson loops, i.e. a consistent power
counting scheme emerges.  Denoting by $B_v$ the heavy baryons, the
correct order is given by the chain
\begin{equation}
\tilde{{\cal L}}_{\rm eff} [U, M , B, B^*] \to
\bar{{\cal L}}_{\rm eff} [U, B] \to {\cal L}_{\rm eff} [U, B_v] \, \, \, ,
\label{chain}
\end{equation}
i.e. the heavy fermion limit has to be taken at the very end.
In Eq.(\ref{chain}), $M$ denotes the various meson resonances and $B^*$
is a generic symbol for any baryon excitation.

The transformation properties of the
Goldstones and the baryons under the chiral group ${\cal G} = SU(3)_L
\times SU(3)_R$ are standard,
\begin{equation}
U \to g_L \, U \, g_R^{-1} \, \, , \quad
B \to h(\phi,g) \, B \, h(\phi,g)^{-1}
\label{ccwz}
\end{equation}
with $(g_L,g_R) \in {\cal G}$ and $h(\phi, g)$ is the so--called
compensator field representing an element of the conserved subgroup
$SU(3)_V$. We also need the square root of $u(\phi) = \sqrt{U(\phi )}$
with the chiral transformation
\begin{equation}
u(\phi ) \to g_L \, u(\phi ) \, h(\phi,g)^{-1} = h(\phi, g)
u(\phi ) g_R^{-1} \,  \, \, .
\label{sqrtu}
\end{equation}
The form of the lowest order meson--baryon
Lagrangian is standard, see e.g. \cite{bkmz}, and the
meson Lagrangian is given in \cite{gl85}. We just need the form of the
chiral covariant derivative $D_\mu$,
\begin{equation}
D_\mu \, B  = \partial_\mu \, B + [ \Gamma_\mu \, , \, B ]
\equiv [D_\mu \, , \, B] \, \, \, ,
\label{cova1}
\end{equation}
with the connection $\Gamma_\mu$,
\begin{equation}
\Gamma_\mu  = \frac{1}{2} \biggl( u^\dagger [\partial_\mu
-i(v_\mu+a_\mu)]u + u [\partial_\mu-i(v_\mu-a_\mu) ]u^\dagger
\biggr) \, \, \, ,
\label{cova2}
\end{equation}
and $v_\mu$ and $a_\mu$ are external vector and axial--vector sources.
The covariant derivative transforms linearly under ${\cal G}$,
\begin{equation}
D_\mu \, B \to h(\phi, g) \, (D_\mu \, B) \, h(\phi , g)^{-1}
\, \, \, .
\label{covg}
\end{equation}
The explicit form of the higher dimension  Goldstone boson--baryon
and meson Lagrangians is not needed here. We stress that the mesonic
LECs show up in ${\cal L}_{\phi}^{(4)}$ whereas  ${\cal L}_{\phi
  B}^{(2,3,4)}$ contain the baryon LECs under discussion here.
In what follows, we will mostly work in the chiral limit. This amounts
to neglecting terms including the external scalar and pseudoscalar
sources, in particular the quark mass matrix. It is straightforward to
generalize our considerations to include such terms.

\subsection{Addition of Spin--1 Fields: Tensor Field Formulation}

First, let us discuss the chiral couplings of spin--1 fields
to the Goldstone bosons in the tensor field formulation. We use the
notation of Refs.\cite{reso,reso1} and only write down the necessary
definitions for our extension to include baryons. Representing massive
spin--1 fields by an antisymmetric tensor is quite common in
e.g. supergravity theories. One of the main advantages in the hadron
sector is that there is no mixing between axials and pseudoscalars in
such an approach. Our basic object is the antisymmetric tensor field
$W_{\mu \nu}$ (which is a generic symbol for any vector or axial--vector),
\begin{equation}
W_{\mu \nu} = - W_{\nu \mu} \, \, \, .
\label{defwmunu}
\end{equation}
This tensor has six degress of freedom. Freezing out the three fields
$W^{ij}$, one reduces this number to three as necessary for massive
spin--1 fields. The free Lagrangian takes the form \cite{reso}
\begin{equation}
{\cal L}_0 = - \frac{1}{2} \partial^\mu W_{\mu \nu}^a \, \partial_\rho
W^{\rho \nu, a} + \frac{1}{4} M_V^2 \, W_{\mu \nu}^a W^{\mu \nu , a}
\, \, \, \, ,
\label{ltfree}
\end{equation}
with $M_V$ the average mass of the spin--1 octet (in the chiral limit)
and the flavor index $a$ runs from 1 to 8.. For non-linearly realized
chiral symmetry, $W_{\mu \nu}$ transforms as any matter field under
the chiral group ${\cal G}$,
\begin{equation}
W_{\mu \nu} \to h(\phi,g) \, W_{\mu \nu} \, h(\phi,g)^{-1} \, \, \, .
\label{ccwzw}
\end{equation}
We point out that other spin--1 field representations like e.g. massive
Yang--Mills fields, have  more complicated transformation properties,
see e.g. \cite{ulfpr,reso1}. Without any assumptions about the nature
of the massive vectors or axials, chiral symmetry leads to Eq.(\ref{ccwzw}).
{}From the Lagrangian, Eq.(\ref{ltfree}), one easily constructs the propagator,
\begin{eqnarray}
& & G_{\mu \nu , \rho \sigma} (x,y) \, \, \, = \, \, \,
< 0 | T \{ W_{\mu \nu} (x) \, , W_{\rho \sigma} (y) \} |0> \, = \nonumber \\
& = &\frac{i}{M_V^2} \, \int \frac{d^4k}{(2\pi)^4} \,
\frac{e^{i \, k \cdot (x-y)}}{M_V^2 -k^2 - i\epsilon} \,
\bigl[ g_{\mu
  \rho} g_{\nu \sigma} (M_V^2 - k^2) + g_{\mu \rho} k_\nu k_\sigma -
g_{\mu \sigma} k_\nu k_\rho \nonumber \\
& & \qquad \qquad \qquad \qquad \qquad \qquad \qquad \qquad \qquad \qquad
- (\mu \leftrightarrow \nu) \, \bigr]
 \, \, \, .
\label{propw}
\end{eqnarray}
The lowest order interaction Lagrangian with Goldstones and external
vector and axial--vector sources  is given by
\begin{equation}
{\cal L}_{\rm int} = \frac{1}{2\sqrt{2}} \, \biggl( F_V \, {\rm Tr} (
W_{\mu \nu} \, f_+^{\mu \nu} ) + i G_V \, {\rm Tr} (W_{\mu \nu} \, [
u^\mu , u^\nu] ) \biggr)
\, \, \, ,
\label{ltint}
\end{equation}
with
\begin{equation}
f_+^{\mu \nu}  = u F_L^{\mu \nu} u^\dagger + u^\dagger F_R^{\mu \nu} u
\, \, , \, \, u_\mu = i u^\dagger \nabla_\mu U u^\dagger = u_\mu^\dagger
\, \, \, ,
\label{fmnu}
\end{equation}
where $F_L^{\mu \nu}$ ($F_R^{\mu \nu}$) are the field strengths
associated to $v_\mu +a _\mu$ ($v_\mu - a _\mu$) and $\nabla_\mu $ is
the chiral covariant derivative acting on the Goldstone bosons. The
two real constants $F_V$ and $G_V$ are not fixed by chiral
symmetry. Their numerical values are $F_V = 154 \,$MeV and $G_V = 69
\,$MeV inferred from $\Gamma (\rho^0 \to e^+ e^-)$ and
$\Gamma (\rho \to \pi \pi)$, in order \cite{reso}. These are all the
ingredients we need from the mesonic sector. For more details, see
Refs.\cite{reso,reso1}.

We now turn to the inclusion of baryons, i.e. the construction of the
lowest order $B\phi W$ couplings. For the resonance exchange processes
to be discussed here, it sufficient to list all terms  to order
${\cal O}(q)$. From these, we can construct all Feynman graphs to
${\cal O}(q^3)$ since we have at least two pions (two $u_\mu$'s of
order $q$) or one field strength tensor of order $q^2$ attached to the
propagating spin--1 field (which also couples to the baryons).
Imposing chiral symmetry, $P$, $C$ and $T$, hermiticity and so on, one finds
(some details are given in Appendix A)
\begin{eqnarray}
{\cal L}_{B\phi W}^{(1)}
& = & R_{D/F} \, {\rm Tr} \, \bigl[ \bar{B} \, \sigma^{\mu \nu} \, (
W_{\mu \nu} \, , \, B) \, \bigr] \nonumber \\
& + & S_{D/F} \, {\rm Tr} \, \bigl[ \bar{B} \, \gamma^\mu \, ( [D^\nu,
W_{\mu \nu}]\, , \, B) \, \bigr] \nonumber \\
& + & T_{D/F} \, {\rm Tr} \, \bigl[ \bar{B} \, \gamma^\mu \, ( [D_\lambda,
W_{\mu \nu}]\, , \, [D^\lambda ,[D^\nu , B]] \, ) \, \bigr]
\nonumber \\
& + & U_{D/F} \, {\rm Tr} \, \bigl[ \bar{B} \, \sigma^{\lambda \nu} \, (
W_{\mu \nu} \, , \, [ D_\lambda , [ D^\mu , B]] \, ) \, \bigr]
\, \, \, ,
\label{Pit}
\end{eqnarray}
\vfill
where we have introduced a compact notation. The symbol $(A,B)$
denotes either the commutator, $(A,B) = [A,B]$, or the anticommutator,
$(A,B) = \{A,B\}$. In the first case, the coupling constant has the
subscript $'F'$, in the second one the subscript $'D'$ applies. To be
specific, the first term in Eq.(\ref{Pit}) thus means
\begin{equation}
R_D \, {\rm Tr} \bigl[ \bar{B} \, \sigma^{\mu \nu} \, \{
W_{\mu \nu} \, , \, B \} \, \bigr] +
R_F \, {\rm Tr} \bigl[ \bar{B} \, \sigma^{\mu \nu} \, [
W_{\mu \nu} \, , \, B ] \, \bigr] \, \, \, .
\label{Pitex}
\end{equation}
The real constants $(R,S,T,U)_{D/F}$ are not fixed by chiral symmetry,
their numerical values will be discussed later. At this stage, we only
assume that they are non--vanishing. From  this Lagrangian, we can
immediately construct the effective action for single spin--1 meson
exchange (we specialize here to the vectors without any loss of generality)
\begin{eqnarray}
S_{\rm eff} &  = & i  \int  d^4x  d^4y
  \biggl\{ R_{D/F} \, {\rm Tr} \, \biggl[ \bar{B}(y) \, \sigma^{\rho \sigma} \,
G_{\mu \nu , \rho \sigma} (x,y) \, ( \, J^{\mu \nu} (x) \, , \, B(y)
\, ) \biggr] \nonumber \\
& + &  S_{D/F} \, {\rm Tr} \, \biggl[ \bar{B}(y) \, \gamma^\rho \,
\partial^\sigma_y G_{\mu \nu , \rho \sigma} (x,y) \, ( \, J^{\mu \nu}
(x) \, , \, B(y) \, ) \biggr] \nonumber \\
& + & T_{D/F} \, {\rm Tr} \, \biggl[ \bar{B}(y) \, \gamma^\rho \,
\partial_{y,\lambda} G_{\mu \nu , \rho \sigma} (x,y) \, ( \, J^{\mu \nu}
(x) \, , \, [ D^\lambda , [D^\sigma ,B(y)]] \, ) \biggr] \nonumber \\
&  + & U_{D/F} \, {\rm Tr} \, \biggl[ \bar{B}(y) \, \sigma^{\lambda \sigma} \,
G_{\mu \nu , \rho \sigma} (x,y) \, ( \, J^{\mu \nu} (x) \, , \,
[D_\lambda , [ D^\rho , B(y) ]] \, ) \biggr] \biggr\} \, \, \, ,
\nonumber \\ & &
\label{sefft}
\end{eqnarray}
where the antisymmetric currents $J^{\mu \nu}= -J^{\nu \mu}$ are defined via
\begin{equation}
J^{\mu \nu} = \frac{1}{2 \sqrt{2} } \, \biggl( F_V \,
f_+^{\mu \nu} + i \, G_V \, [u^\mu , u^\nu] \, \biggr) \,
\, \, . \label{Jt}
\end{equation}
The currents $J^{\mu \nu}$ obviously are of chiral dimension two. With
the effective action Eq.(\ref{sefft}) at hand, we are now in the
position of calculating the desired matrix--elements involving single
vector meson exchange contributions to e.g. the nucleon electromagnetic
form factors or elastic pion--nucleon scattering (in the tensor field
formulation). Notice that for these processes it is sufficient to keep
the partial derivatives acting on the $G_{\mu \nu , \rho \sigma}$
in Eq.(\ref{sefft}) and not the chiral covariant ones. The generalization is
obvious and not treated here.

\subsection{Addition of Spin--1 Fields: Vector Field Formulation}

We now consider the three vector $V_\mu$ representing the three
degrees of freedom of a massive spin--1  vector field (the
generalization to axials is straightforward). We do not assign any
particular microscopic dynamics to this vector field, like it is e.g
the case in the hidden symmetry \cite{bky} or the massive Yang--Mills
approaches \cite{ulfpr}. $V_\mu$ is assumed to have the standard
transformation properties of any matter field under the non--linearly
realized chiral symmetry,
\begin{equation}
V_{\mu} \to h(\phi,g) \, V_{\mu} \, h(\phi,g)^{-1} \, \, \, .
\label{ccwzv}
\end{equation}
The corresponding free--field Lagrangian reads
\begin{equation}
{\cal L}_0 = - \frac{1}{4} {\rm Tr} \, \biggl( V_{\mu \nu} \,
V^{\mu \nu}  -  2 \,  M_V^2 \, V_{\mu} V^{\mu} \, \biggr)\, \, \, \, ,
\label{lvfree}
\end{equation}
with the (abelian) field strength tensor
\begin{equation}
V_{\mu \nu} = D_\mu \, V_\nu - D_\nu \, V_\mu
\, \, \, \, .
\label{defvmunu}
\end{equation}
The relation to the massive Yang--Mills or hidden symmetry approach is
discussed in Ref.\cite{reso1}. The propagator is readily evaluated,
\begin{eqnarray}
G_{\mu \rho} (x,y) &=& <0| \,T \bigl\{ \,V_\mu (x) \, , \,  V_\rho (y)
\, \bigr\} \, |0 > \nonumber \\
&=& -i  \int  \frac{d^4k}{(2\pi )^4} \, {\rm
  e}^{-i k\cdot (x-y)} \, \frac{g_{\mu \rho} - k_\mu k_\rho/M_V^2}{k^2
  - M_V^2 + i \epsilon} \, \, \, \, .
\label{propv}
\end{eqnarray}
For the effective action discussed below, we also need the following
three-- and four--point functions
\begin{eqnarray}
-i \, \Delta_{\mu \nu ,  \rho} (x,y) &=& <0| \,T \bigl\{ \,V_{\mu \nu}  (x)
\, , \,  V_\rho (y)
\, \bigr\} \, |0 > \, \, \, \, , \nonumber \\
-i \, \Delta_{\mu \nu ,  \rho \sigma} (x,y) &=&
<0| \,T \bigl\{ \,V_{\mu \nu}  (x)  \, , \,  V_{\rho \sigma} (y)
\, \bigr\} \, |0 > \, \, \, \, , \nonumber \\
-i \, \Delta_{\mu \nu ,  \rho \sigma}^\dagger (x,y) &=&
<0| \,T \bigl\{ \,\bar{V}_{\mu \nu}  (x)  \, , \,  \bar{V}_{\rho \sigma} (y)
\, \bigr\} \, |0 >  \, \, \, \, ,
\label{def34pt}
\end{eqnarray}
with $\bar{V}_{\mu \nu} = D_\mu \, V_\nu + D_\nu \, V_\mu$.
These three- and four--point functions
 are related to the tensor field propagator, Eq.(\ref{propw}),
and its derivative via
\begin{eqnarray}
& &\frac{1}{i} \, \partial_y^\sigma \, G_{\mu \nu , \rho \sigma} ( x,y)
 = - \Delta_{\mu \nu ,  \rho} (x,y)
\, \, \, \, , \nonumber  \\
& &\frac{1}{i} \, G_{\mu \nu , \rho \sigma} ( x,y)  =
-\frac{1}{M_V^2} \, \Delta_{\mu \nu ,  \rho \sigma} (x,y)
+  \frac{1}{M_V^2} \, \bigl( \,g_{\mu \rho} \, g_{\nu \sigma}
 - g_{\mu \sigma} \, g_{\nu \rho} \, \bigr) \, \delta^{(4)}(x-y)
\nonumber \, \, \, \, ,\\ & &
\label{relat}
\end{eqnarray}
where the contact terms in the second relation will be of importance
later on (see also Ref.\cite{reso1}). The interaction Lagrangian of
the vectors with pions and external vector and axial--vector sources
takes the form
\begin{equation}
{\cal L}_{\rm int} = -\frac{1}{2\sqrt{2}} \, \biggl( f_V \, {\rm Tr} (
V_{\mu \nu} \, f_+^{\mu \nu} ) + i g_V \, {\rm Tr} (V_{\mu \nu} \, [
u^\mu , u^\nu] ) \biggr)
\, \, \, ,
\label{lvint}
\end{equation}
and equivalence to the tensor field formulation for the spin--1 pole
contributions (which are of order $q^6$) is achieved by setting \cite{reso1}
\begin{equation}
f_V = \frac{F_V}{M_V} \, \, , \quad g_V = \frac{G_V}{M_V} \, \, .
\label{equivmes}
\end{equation}

Including the baryon octet $B$ leads to the following terms at order
${\cal O}(q)$ for the $B\phi V$ interactions (see again appendix A for
some details on the construction of these terms)
\begin{eqnarray}
{\cal L}_{B\phi V}^{(1)}
& = & K_{D/F} \, {\rm Tr} \, \bigl[ \bar{B} \, \sigma^{\mu \nu} \, (
V_{\mu \nu} \, , \, B) \, \bigr] \nonumber \\
& + & G_{D/F} \, {\rm Tr} \, \bigl[ \bar{B} \, \gamma^\mu \, (
V_\mu\, , \, B) \, \bigr] \nonumber \\
& + & L_{D/F} \, {\rm Tr} \, \bigl[ \bar{B} \, \sigma^{\mu \nu} \, (
V_{\mu \lambda} \, , \, [D^\lambda ,[D_\nu , B]] \, ) \, \bigr]
\nonumber \\
& + & H_{D/F} \, {\rm Tr} \, \bigl[ \bar{B} \, \sigma^{\mu \nu} \, (
{\bar V}_{\mu \lambda} \, , \, [ D^\lambda , [ D_\nu , B]] \, ) \, \bigr]
\, \, \, ,
\label{Piv}
\end{eqnarray}
The first term in Eq.(\ref{Piv}) when contracted with two $u_\mu 's$
or a  field
strength tensor contributes to order $q^3$ while the other terms only
start one order higher. However, for the equivalence of the spin--1
pole contributions they have to be retained, as shown in the next
subsection (compare to  the spin--1 pole terms in the meson sector which are
of order $q^6$). The fourth term in Eq.(\ref{Piv}) is of relevance for
some additional  $q^4$ terms. Because it includes the symmetric tensor
$\bar{V}_{\mu \nu} = \bar{V}_{\nu \mu}$, it apparently has no
counterparts in the tensor field formulation, where only the antisymmetric
tensor $W_{\mu \nu}$ appears. That there is, however, no inequivalence
is shown in appendix A.
The real constants $(K,G,H,L)_{D/F}$ are not fixed by chiral symmetry,
we only assume that they are non--vanishing. The corresponding
effective action takes the form
\begin{eqnarray}
S_{\rm eff} &  = & - \int \, d^4x \, d^4y
  \biggl\{ K_{D/F} \, {\rm Tr} \, \biggl[ \bar{B}(y) \,
  \sigma^{\lambda \rho} \,
\Delta_{\mu \nu , \lambda \rho} (x,y) \, ( \, J^{\mu \nu} (x) \, , \, B(y)
\, ) \biggr] \nonumber \\
& + &  G_{D/F} \, {\rm Tr} \, \biggl[ \bar{B}(y) \, \gamma^\rho \,
\Delta_{\mu \nu , \rho } (x,y) \, ( \, J^{\mu \nu}
(x) \, , \, B(y) \, ) \biggr] \nonumber \\
& + & L_{D/F} \, {\rm Tr} \, \biggl[ \bar{B}(y) \, \sigma^{\lambda \rho} \,
\Delta_{\mu \nu , \lambda \sigma} (x,y) \, ( \, J^{\mu \nu}
(x) \, , \, [ D_\rho , [D^\sigma ,B(y)]] \, ) \biggr] \nonumber \\
&  + & H_{D/F} \, {\rm Tr} \, \biggl[ \bar{B}(y) \, \sigma^{\lambda \rho} \,
\Delta_{\mu \nu , \lambda \sigma}^\dagger (x,y) \,
( \, J^{\mu \nu} (x) \, , \,
[D^\sigma , [ D_\rho , B(y) ]] \, ) \biggr] \biggr\} \, \, \, ,
\nonumber \\ & &
\label{seffv}
\end{eqnarray}
and the antisymmetric currents $J_{\mu \nu}$ are expressed in terms of
the constants $f_V$ and $g_V$. Notice that terms of the type
Tr$(\bar{B} D_\mu V^\mu B)$ do not contribute to the effective action
as shown in appendix A. This completes the formalism necessary
to proof the equivalence between the tensor and vector field approaches.

\subsection{Equivalence for Spin--1 Pole Contributions}

We are now in the position to establish in steps the equivalenve
between the effective Lagrangians ${\cal L}_{B\phi W}^{(1)}$ and
${\cal L}_{B\phi V}^{(1)}$. Comparing the effective actions,
Eqs.(\ref{sefft},\ref{seffv}), and making use of the first relation in
Eq.(\ref{relat}), one establishes the complete equivalence of the
terms at order $q^3$
\begin{equation}
 S_{D/F} \, {\rm Tr} \, \bigl[ \bar{B} \, \gamma^\mu \, ( [D^\nu,
W_{\mu \nu}]\, , \, B) \, \bigr]
\quad {\rm and } \quad
 G_{D/F} \, {\rm Tr} \, \bigl[ \bar{B} \, \gamma^\mu \, (
V_\mu\, , \, B) \, \bigr]
\end{equation}
if one identifies the coupling constants as follows
\begin{equation}
 G_{D/F} = - M_V \,  S_{D/F} \,  \, \, .
\label{equiv1}
\end{equation}
Furthermore, the spin--1 pole contributions of the two terms
\begin{equation}
R_{D/F} \, {\rm Tr} \, \bigl[ \bar{B} \, \sigma^{\mu \nu} \, (
W_{\mu \nu} \, , \, B) \, \bigr] \quad {\rm and} \quad
 K_{D/F} \, {\rm Tr} \, \bigl[ \bar{B} \, \sigma^{\mu \nu} \, (
V_{\mu \nu} \, , \, B) \, \bigr]
\end{equation}
are equivalent at order $q^4$ if
\begin{equation}
 R_{D/F} = - M_V \,  K_{D/F} \,  \, \, .
\label{equiv2}
\end{equation}
Similarly, one derives
\begin{equation}
 U_{D/F} =  M_V \, (\, H_{D/F} - L_{D/F} \,)  \, \, ,
\label{equiv2a}
\end{equation}
making use of Eq.(\ref{a16}), see appendix A. Using that
decomposition, there is another term proportional to $H_{D/F}$,
$H_{D/F} {\rm Tr}[\bar{B} \sigma^{\mu \nu} (D_\mu D^\rho W_{\rho
  \lambda} \, , \, D^\lambda D_\nu B ) ] $ which contributes to the processes
under consideration at order $q^4$. These higher
order terms we do not consider in their full generality.
There is, however, also a  difference at order $q^2$ due to the contact
term in $G_{\mu \nu, \rho \sigma}$, compare the second relation in
Eq.(\ref{relat}). To pin this down, we have to compare some physical
observables as explained below. It is important to stress the formal
equivalence between the effective actions, Eq.(\ref{sefft}) and
Eq.(\ref{seffv}), which follows by inspection if one sets
\begin{equation}
V_\mu = -\frac{1}{M_V} \, \partial^\nu \, W_{\nu \mu}  \, \, \, .
\label{formeq}
\end{equation}
This is a consequence of the first identity in Eq.(\ref{relat}) because up
to some contact terms, it means that $V_{\mu \nu} \simeq W_{\mu \nu}$. One
can show that in
general, the differences between the propagators in the vector and the
tensor field formulation lead to different local effective
actions. Physical arguments are needed to pin down the coefficients of
these polynomials. What remains to be done is therefore to enumerate the
local terms that differ in the vector and tensor field formulation and to
determine their coefficients.


\section{Electromagnetic Form Factors for the Baryon Octet}

In this section, we want to establish the equivalence between the
tensor and the vector field formulation by calculating the
vector meson contribution to the electromagnetic form factors of the
baryon octet. This is analogous to the pion form factor calculation
performed in Ref.\cite{reso1}. First, we will give some basic definitions
and then establish the equivalence.

\subsection{Basic Definitions}

Consider an electromagnetic external vector source,
\begin{equation}
v_\mu = - e \, A_\mu \, Q \, \, \, ,
\label{vem}
\end{equation}
where $e > 0$ is the unit charge, $A_\mu$ denotes the photon field and
$Q= (1/3) \,$ diag$(2,-1,-1)$ is the quark charge matrix. We also
have for the chiral connection $\Gamma_\mu = -i v_\mu +{\cal
  O}(\phi^2)$. In the tensor formulation, the antisymmetric currents
$J_{\mu \nu}$ simplify to
\begin{equation}
J^{\mu \nu} = \frac{1}{2 \sqrt{2}} \, F_V \, f_+^{\mu \nu} =
- \frac{e}{\sqrt{2}} \, F_V \, Q \, (\partial^\mu A^\nu -
\partial^\nu A^\mu ) \, \, ,
\label{jmunuem}
\end{equation}
and similarly for the vector formulation, substituing $F_V$ by $f_V$.
The transition matrix element for the electromagentic current follows
by differentiation of the effective action,
\begin{equation}
<B(p') | \, J_\mu^a (0) \, | B(p)> = \,
\,  <B(p') | \,  \frac{\delta S_{\rm eff}}{\delta v^\mu_a (0)}
\, | B(p)>
 \, \, \, .
\label{jmuseff}
\end{equation}
For the following discussion, it is most convenient to express the
current matrix element in terms of {\it four} form
factors,
\begin{eqnarray}
& & <B(p') | \, J_\mu^a \, | B(p)> \, \, = \, \,
<B(p') | \, \bar{q} \, \gamma_\mu \lambda^a \, q \,
| B(p)>\nonumber \\
& = & {\rm Tr} \, \biggl( \bar{u} (p') \bigl[ \, \gamma_\mu \, F_1^+ (t) +
\frac{i \, \sigma_{\mu \nu}k^\nu}{2m} \, F_2^+ (t) \bigr] \{ \lambda^a
, u(p) \} \biggr) \nonumber \\
& + & {\rm Tr} \, \biggl( \bar{u} (p') \bigl[ \, \gamma_\mu \, F_1^- (t) +
\frac{i \, \sigma_{\mu \nu}k^\nu}{2m} \, F_2^- (t) \bigr] [ \lambda^a
, u(p) ] \biggr)  \, \, ,
\label{ffdef}
\end{eqnarray}
with $u(p) = u^a (p) \lambda^a$, $\bar{u}(p) = \bar{u}^a (p)
{\lambda^a}^\dagger$ and $t = (p'-p)^2$ is the invariant momentum
transfer squared. The more commonly used Dirac ($F_1$) and Pauli
($F_2$) form factors are simple linear combinations of the $F_1^\pm$
and $F_2^\pm$, respectively, for any baryon state $B(p)$. For example,
the Pauli form factor of the proton is given by
$F_2^p (t)= F_2^- (t) + (1/3) F_2^+ (t)$.

\subsection{Vector Meson Contribution to the Electromagnetic Form Factors}

We now calculate the vector meson contribution to the form factors
$F_{1,2}^\pm (t)$. A full chiral calculation would also involve the
one loop contributions for the isovector form factors, these are e.g.
given in Refs.\cite{gss,bkkm}. For our purpose, these terms are of no
relevance and are thus neglected here.

Consider first the tensor field formulation. From the effective
action, Eq.(\ref{sefft}), it is straightforward to get the one--baryon
transition matrix elements. To order $q^3$ in the effective action,
i.e. to order $q^2$ in the form factors, these read
\begin{eqnarray}
F_1^+ (t) &=& F_V \, \frac{t}{t-M_V^2} \, \biggl\{ \sqrt{2} S_D -
\frac{1}{\sqrt{2}} \, m \, U_D \, \biggr\}
- \frac{1}{2\sqrt{2}} \, F_V \, T_D \, \frac{t^2}{t-M_V^2}
\, \, , \nonumber \\
F_2^+ (t) &=& F_V \, 4 \sqrt{2} \, m \, R_D \, \frac{1}{t-M_V^2} \,
 \, \, , \nonumber \\
F_1^- (t) &=& 1 + F_V \, \frac{t}{t-M_V^2} \, \biggl\{ \sqrt{2} S_F -
\frac{1}{\sqrt{2}} \, m \, U_F \, \biggr\}
- \frac{1}{2\sqrt{2}} \, F_V \, T_F \, \frac{t^2}{t-M_V^2}
\, \, , \nonumber \\
F_2^- (t) &=& F_V \, 4 \sqrt{2} \, m \, R_F \, \frac{1}{t-M_V^2} \,
 \, \, ,
\label{fft}
\end{eqnarray}
with $m$ the average octet mass (in the chiral
limit).\footnote{Throughout, we do not differentiate between the
physical (average) octet baryon mass and its
chiral limit value.}  We notice that
the terms proportional to $T_{D/F}$ only start to contribute at order
$q^4$. We remark that in the tensor field approach, $F_1^\pm (0) =0$
and $F_2^\pm (0) \neq 0$. This in particular means that there is a
finite vector meson induced LEC
contributing to the anomalous magnetic moment $\kappa_B$.
The numerical values of the LECs appearing in Eq.(\ref{fft}) could be
fixed from the measured baryon elecromagnetic radii and magnetic
moments. We do not attempt this determination here since it is not
necessary for the following discussions.
A similar calculation for the vector field formulation to the same
order gives
\begin{eqnarray}
F_1^+ (t) &=& - \sqrt{2} \, f_V \, G_D \, \frac{t}{t-M_V^2}
+ \frac{m \, f_V}{\sqrt{2}} \, \biggl( \, L_D
-   H_D \, \biggr) \, \frac{t^2}{t-M_V^2}
\, \, , \nonumber \\
F_2^+ (t) &=& - 4 \sqrt{2} \, m \, f_V \, K_D \, \frac{t}{t-M_V^2} \,
+ \sqrt{2} \, m  \, f_V \, H_D \, \frac{t^2}{t-M_V^2}
 \, \, , \nonumber \\
F_1^- (t) &=& 1 - \sqrt{2}\, f_V \, G_F \, \frac{t}{t-M_V^2}
+ \frac{m \, f_V}{\sqrt{2}} \, \biggl( \, L_F
-  H_F \, \biggr) \,  \frac{t^2}{t-M_V^2}
\, \, , \nonumber \\
F_2^- (t) &=& - 4 \sqrt{2} \, m \, f_V \, K_F \, \frac{1}{t-M_V^2} \,
+ \sqrt{2} \, m  \, f_V \, H_F \, \frac{t^2}{t-M_V^2}
 \, \, ,
\label{ffv}
\end{eqnarray}
where in striking contrast to the tensor formulation, Eq.(\ref{fft}),
we now find $F_2^{\pm} (0) =0$. This would imply that there is no
vector meson induced contribution to the anomalous magnetic moment of
any of the octet baryons. Comparing Eq.(\ref{fft}) with
Eq.(\ref{ffv}), we have equivalence if the LECs obey
\begin{eqnarray}
 M_V \, K_{D/F}& = & - R_{D/F}  \, , \, \, G_{D/F} = - S_{D/F} \,  M_V
\, , \nonumber \\
U_{D/F} & = & M_V \,(H_{D/F}- L_{D/F})   \, \, \, .
\label{equiv3}
\end{eqnarray}
These are, of course, the already known spin--1 pole
equivalences. Clearly, one needs a finite vector meson induced LEC to
the anomalous magnetic moments as it is the case in the tensor field
formulation. The situation is similar to the one concerning the
electromagnetic form factor of the pion \cite{reso1}, where in the
vector representation the radius vanishes while it is finite and close
to the successful vector meson dominance value in the tensor field
approach. A recent dispersion--theoretical study of the nucleon form
factors indeed substantiates the relevance of vector mesons
in the description at low and intermediate momentum
transfer \cite{mmd}. The differences for the form factors calculated in
vector and tensor formulation are indeed due to the local contact
terms in which $G_{\mu \nu , \rho\sigma}$ and $\Delta_{\mu \nu ,
  \rho\sigma}$ differ. To achieve complete equivalence, one therefore has
to add in the vector field approach a local dimension two
operator of the type
\begin{equation}
-R_{D/F} \, {\rm Tr} \, \biggl[ \bar{B}(x) \, \sigma^{\rho \sigma} \,
\frac{1}{M_V^2} (\, g_{\mu \sigma} \, g_{\nu \rho} - g_{\mu \rho} \,
g_{\nu \sigma} \, ) \, ( J^{\mu \nu}(x) \, ,\, B(x) \, ) \, \biggr] \,
\, .
\label{localff}
\end{equation}
Then, the vector meson contributions to the baryon electromagnetic form
factors are indentical in the tensor and vector field formulation to
order ${\cal O}(q^3)$. Higher order terms, which will modify some of the
$t^2$--terms, are not considered here (they could be treated in a
similar fashion).  Having achieved this equivalence, we now come
back to the problem of the vector meson contribution to the
isovector pion--nucleon amplitude in forward direction.

\vfill

\section{Vector Meson Contribution to Pion--Nucleon Scattering }

Consider the vector meson contribution to the elastic Goldstone boson--baryon
scattering amplitude, in particular  $T( \pi(q) + N(p) \to \pi(q') +
N(p') )$, cf. Fig.1b. We will evaluate this for the vector and the
tensor field formulations. In doing so, we make use of momentum
conservation, $p+q = p' + q'$, the fact that the baryons are
on--shell,
\begin{equation}
( p\!\!\!/ -m) \,u(p) = \, 0 \,=  \,
\bar{u}(p') \,  ( p\!\!\!/ ' -m) \, \, \, ,
\end{equation}
and also the Gordon decomposition,
\begin{equation}
i \, \bar{u}(p') \,  \sigma^{\mu \nu} \, ( p' -p)_\nu \, u(p) =
\, \bar{u}(p') \,  \bigl( \,2 m \gamma^\mu  - ( p' +p)^\mu \,\bigr) \, u(p)
\, \, .
\end{equation}
Furthermore, we introduce the short-hand notation
\begin{equation}
{\cal X}_{D/F} = X_{D/F} \, {\rm Tr} \,
\biggl( \, ( \lambda^A , {\lambda^B}^\dagger
) \, [ \lambda^C , {\lambda^D}^\dagger ] \, \biggr) \, \, ,
\label{nota}
\end{equation}
where $X$ stands for any of the LECs $G,H, \ldots ,S,T,U$
appearing in Eqs.(\ref{Pit},\ref{Piv}) and $(A,B)$ as defined
below Eq.(\ref{Pit}).

For the tensor field formulation, the calculation of the process
$\phi + B \to W \to \phi' +B'$ results in
\begin{eqnarray}
& &\frac{-i \sqrt{2} G_V}{F^2_\pi  (M_V^2-t)} \biggl\{ \bigl[ - 2 m
( q\!\!\!/ + q\!\!\!/ ') + ( q+q') \cdot (p+p') \bigr]  \, \bigl[ {\cal R}_D +
{\cal R}_F \, \bigr] \nonumber \\
& + & \frac{t}{2}  \, (q\!\!\!/ + q\!\!\!/ ')\, \bigl[ {\cal S}_D +
{\cal S}_F \, \bigr] + \frac{t}{4} \, m \, ( q\!\!\!/ + q\!\!\!/ ')
\bigl[ {\cal U}_D + {\cal U}_F \, \bigr] \nonumber \\
& - &  \frac{t^2}{8}  (q\!\!\!/ + q\!\!\!/ ' ) \,
\bigl[ {\cal T}_D + {\cal T}_F \, \bigr] \, \biggr\} \, \, \, \, .
\label{pint}
\end{eqnarray}
We note that in forward direction, $p' \to p$, i.e. $t \to 0$, this
expression vanishes since
\begin{equation}
q \cdot (p+p') \to p \cdot q + q \cdot p = \{ p\!\!\!/ , q\!\!\!/ \}
= 2 m \, q\!\!\!/ \, \, .
\end{equation}
Therefore, the first corretion to the LET for $T^- (\nu) / \nu$ in
forward direction is suppressed by one power in small momenta, i.e.
the Weinberg term is not modified. This should be compared with the
result in the vector field formulation,
\begin{eqnarray}
\frac{i 2 \sqrt{2} g_V}{F^2_\pi  (M_V^2-t)} \, t
 & \biggl\{ & \bigl[ - 2 m
(q\!\!\!/ ' + q\!\!\!/ ) +  (q'+q) \cdot (p' +p) \bigr] \,
\bigl[ {\cal K}_D + {\cal K}_F \, \bigr] \nonumber \\
& + & \frac{1}{2} \, ( q\!\!\!/ + q\!\!\!/ ') \, \bigl[ {\cal G}_D +
{\cal G}_F \, \bigr]
 +  \frac{m}{4} \, t \, ( q\!\!\!/ + q\!\!\!/ ') \, \bigl[ {\cal L}_D +
{\cal L}_F \, \bigr] \nonumber \\
& + & \frac{t}{2} \bigl[ -\frac{1}{2} \, m \, (q\!\!\!/ ' + q\!\!\!/ )
+ (p' + p) (q' + q) \bigr] \,\bigl[ {\cal H}_D + {\cal H}_F \, \bigr]
\, \biggr\}  \, \, \, ,
\label{pinv}
\end{eqnarray}
which vanishes identically in forward direction. For the equivalence
to hold, one therefore has to add a local term to the
effective Lagrangian ${\cal L}^{(1)}_{B\phi V}$ which is of the form
as given in Eq.(\ref{localff}) but with the $J_{\mu \nu}$ piece
proportional to $[u_\mu , u_\nu]$, i.e. two pions.
Again, the Weinberg term is not modified but one rather gets some
higher order corrections. In summary, we can say that if one couples
vector mesons to the effective pion--nucleon Lagrangian
solely based on non--linearly realized chiral symmetry, the Weinberg
term is already contained in the covariant derivative acting on the
nucleon fields and the vector mesons simply provide part of the
corrections at next--to--leading order. This is the most natural
framework since these spin--1 fields are not to be considered as
effective degrees of freedom but rather provide certain contributions
to some low--energy constants via resonance saturation, i.e. they are
integrated out from the effective field theory.  Of course, it
is still possible to treat the vector mesons say as massive
Yang--Mills particles and get the correct result. This is discussed in
some detail in section VII of Weinberg's seminal paper \cite{weinnon}
on non--linear realizations of chiral symmetry and we refer the reader
for details to that paper.


\section{Loops with massive particles}

Up to now, we have only considered tree graphs. However, it is
well--known that in the relativistic formulation of baryon CHPT, loops
can contribute below the chiral dimension one would expect (as
detailed by Gasser et al. in Ref.\cite{gss}). Therefore, for the
single vector-exchange graphs we consider, we have to be more careful.
These general diagrams for Goldstone boson--baryon scattering (in
forward direction) or the baryon electromagnetic form factors are
shown in Fig.2. The shaded areas denote the whole class of multi--loop
diagrams. To be precise, these are Goldstone boson loops.
The blob on the right side does only contain Goldstones and
external sources and thus there is a consistent power counting, i.e
the $n$--loop insertions on that vertex are obviously suppressed by
$2n$  powers in the small momentum compared to the leading tree graph. To
leading order, we thus only need to consider the tree level  couplings
of the massive spin--1 fields to the pions\footnote{We generically
  denote the Goldstone bosons as pions.} or the photon. Matters are
different for the loops contained in the left blob in Fig.2. since
there is no more one--to--one correspondence between the loop
expansion and the expansion in small momenta and meson masses.
\begin{figure}[t]
\hskip 2in
\epsfysize=2in
\epsffile{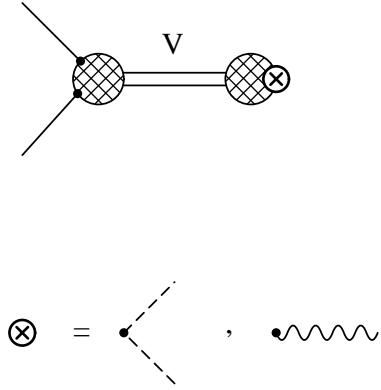}
\caption{\label{fig2} The general single vector meson exchange graph
attached to baryons (straight lines) and pions (dashed lines)  or
external sources (wiggly lines) like e.g. the photon. The shaded areas
denote the loop expansion of the respective vertices. Notice that the
circle-cross is completely embedded in the shaded area on the right.}
\end{figure}
There is no a priori reason why multi--loop graphs can not even modify
the lowest order results, see also ref.\cite{gss}.
We will now show that in the tensor and in the vector
formulation, these type of graphs only contribute to subleading
orders, i.e. they do neither modify the Weinberg term nor the leading
contributions to the anomalous magnetic moments of the octet baryons.
This is intimately related to the form of the $V\phi \phi$ and $V_\mu
A^\mu$ couplings.

Consider first the vector formulation. The two--pion--vector vertex
is proportional to
\begin{equation}
(p'-p)_\nu \, (k+p'-p)^\nu \, k^\mu
\end{equation}
\begin{figure}[b]
\hskip 1.7in
\epsfysize=1.5in
\epsffile{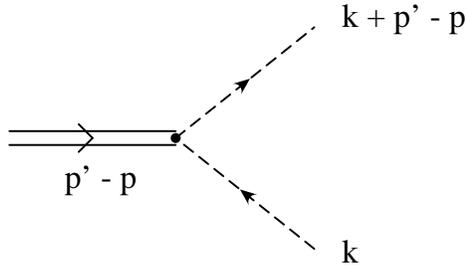}
\caption{\label{fig3} Momentum flow in  the $V\phi \phi$--vertex.}
\end{figure}
where the vector carries momentum $p'-p$ and the in--coming
(out--going) pion $k$ ($k+p'-p$), see Fig.3. This obviously vanishes in forward
direction, $p' \to p$. Similarly, the photon--vector meson vertex is
proportional to $(p'-p)_\nu (p' -p)_\mu$, which is at least of order
$q^2$ and thus can not contribute to the leading order anomalous
magnetic moment term.  In the tensor field formulation, the
vector--two--pion vertex is proprtional to
\begin{equation}
k_\mu \, (k+p'-p)_\nu = k_\mu \, (p' - p)_\nu \, \, \, ,
\end{equation}
for the same routing of momenta as above and we have made use of the
antisymmetry of the tensor field propagator under $\mu \leftrightarrow
\nu$ interchange. Thus, there is no contribution to the Weinberg term
but only to subleading corrections. The eventual differences appearing
in these subleading orders can be absorbed by proper redefinitions of
the LECs $(G,H \ldots, T,U)_{D/F}$.


\section{Short Summary}

In this paper, we have considered the coupling of massive spin--1
fields to the effective field theory of the Goldstone bosons and
the ground state baryon octet. The corresponding Lagrangian is
parametrized beyond
leading order in terms of so--called low--energy constants. These can
be estimated by $s$--channel baryonic  and $t$--channel mesonic
excitations. For the latter type of contributions, massive
spin--1 fields (in particular vector mesons) play an important role.
If one does not assume any particular underlying dynamics for these
fields but only their transformation properties under non--linearly
realized chiral symmetry, one is left with two possibilities, namely
to represent the three degrees of freedom by a conventional vector or
a properly constrained antisymmetric tensor field. One expects these
two representations to be equivalent. That is indeed what we have shown
here. For that, we have constructed the effective action of the
Goldstone boson--baryon-spin--1 system for both types of spin--1
representations. These in general differ by local (contact) terms with
a priori undetermined coefficients.
The study of the baryon electromagnetic form factors allowed us to pin
down these local terms, which have to be added in the vector field
formulation, and to determine their corresponding coefficients. We also
clarified the situation concerning the so--called Weinberg term of
relevance to low energy pion--nucleon scattering. In
the approach discussed here, it arises from the chiral covariant
derivative acting on the baryons and not from vector meson
exchange. The latter type of process contributes at next--to--leading
order, as specifed above. In summary, the situation is very similar to
the one in the meson sector were some local contact terms of dimension
four have to be added in case of the vector field approach to achieve
e.g. a satisfactory description of the pion electromagnetic form
factor at next--to--leading order \cite{reso1} (or, to be precise, the
electromagnetic radius of the pion). Finally, we have briefly
discussed some problems related to loops in the presence of massive
particles. These, however, do not alter any of our conclusions.

\vfill


\appendix

\section{Construction Principles}
\def\theequation{\Alph{section}.\arabic{equation}}
\setcounter{equation}{0}
\label{appa}

Here, we collect a few useful formulae which are needed to construct
the effective Lagrangians ${\cal L}^{(1)}_{B\phi W}$ and
${\cal L}^{(1)}_{B\phi V}$. To the order we are working, we can make
use of the  equation of motion for the baryon fields derived from
${\cal L}^{(1)}_{B\phi}$
\begin{equation}
i\gamma^\mu [D_\mu,B] - mB + \frac{1}{2} D \gamma^\mu \gamma_5\{u_\mu, B\}
+ \frac{1}{2} F \gamma^\mu \gamma_5 [u_\mu, B]  = 0 \, \, .
\label{eom}
\end{equation}
In fact, only the chiral--covariant kinetic and the mass terms will be
of relevance to the order we are working. It is worth to stress that
we only consider processes with at most two external pions or one
external vector source, like e.g. single vector--meson exchange in
Goldstone boson--baryon scattering or $t$--channel vector meson
exciations contributing to the baryon electromagnetic form factors. It
is straightforward to generalize our considerations to more
complicated processes. In what follows, we  frequently  employ our
short--hand notation $(A,B)$ meaning either $[A,B]$ or $\{A,B\}$.

Consider first the tensor
field formulation. We want to construct terms which contain one
antisymmetric tensor and at most two covariant derivatives since in the
low energy domain
\begin{equation}
[D_\mu , [D_\nu , W_{\lambda \kappa}]] = {\cal O}(q^2) \, \, \, .
\end{equation}
Because of the antisymmetry of $W_{\mu \nu}$, many terms vanish. We
give one example
\begin{equation}
{\rm Tr} \, \biggl[ \bar{B} \biggl( [D^\mu , [D^\nu , W_{\mu \nu}] \, ,
\, B \biggr) \biggr] \sim {\rm Tr} \, \biggl[ \bar{B} \biggl( [u^\mu,
u^\nu] , W_{\mu \nu}] \, , \, B \biggr) \biggr] = 0 \, \, \, ,
\label{antis}
\end{equation}
for the processes under consideration. The term on the
right--hand--side of Eq.(\ref{antis}) gives rise to e.g. a
vector-two-pion-baryon seagull term and is thus of no relevance here.
The discrete symmetries $P$, $C$ and $T$ as well as hermiticity and
Lorentz and chiral invariance strongly constrain the possible terms.
Consider e.g. charge conjugation. We have
\begin{equation}
C: \quad W_{\mu \nu}^c = -W_{\mu \nu}^T \, \, , \,
B^c = C \, \bar{B}^T \, \, , \, \bar{B}^c = B^T \, C \, \, , \,
D_\mu^c = -D_\mu^T \, \, ,
\label{chargec}
\end{equation}
and the standard  transformation properties for the Dirac matrices.
This eliminates for example a term like
\begin{equation}
{\rm Tr} \, \biggl[ \bar{B} \gamma^\mu \biggl( D^2 W \, ,
\, D^3B \biggr) \biggr]^c \simeq -{\rm Tr} \, \biggl[ \bar{B}
\gamma^\mu \biggl(
(D^2 , W) \, , \, D^3 B \biggr) \biggr]  \, \, \, ,
\end{equation}
in symbolic notation (i.e. dropping all unnecessary indices).
The '$\simeq$' means that we have dropped terms of higher order which
arise when one moves the covariant derivatives around. Under parity
transformations
\begin{eqnarray}
P: & &\quad P \,W_{\mu \nu}(x) \, P^\dagger
= W^{\mu \nu}(\tilde{x}) \, \, , \,
P \, D_\mu (x) \, P^\dagger = D^\mu (\tilde{x}) \, \, , \nonumber \\
& & \quad P \, B(x) \,P^\dagger = \gamma_0 \, B (\tilde{x}) \, \, , \,
P \, \bar{B}(x) \,P^\dagger =  \bar{B} (\tilde{x}) \, \gamma_0 \,\, \, ,
\label{parity}
\end{eqnarray}
with $\tilde{x}^\mu = x_\mu$. We also note that while $W_{\mu \nu}$ is
hermitean, $W_{\mu \nu}^\dagger = W_{\mu \nu}$, the covariant
derivative is anti-hermitean, $D_\mu^\dagger = - D_\mu$. With these
tools at hand, one readily constructs the effective Lagrangian
Eq.(\ref{Pit}).

For the vector field formulation, one can proceed similarly. Here,
however, it is helpful to remember the standard $\rho$--meson--nucleon
vector and tensor couplings,
\begin{equation}
{\cal L}_{\rho N} = \frac{g}{2} \, \bar{\Psi}_N \, \biggl[ \gamma^\mu \,
\vec{\rho}_\mu \cdot \vec{\tau} - \frac{\kappa_\rho}{2m} \,
\sigma^{\mu \nu} \, \partial_\nu \, \vec{\rho}_\mu \cdot \vec{\tau}
\biggr] \, \Psi_N \, \, \, .
\end{equation}
These terms can easily be transcribed into  SU(3) vector meson-baryon
couplings (in the vector field formulation) of the type
\begin{equation}
{\rm Tr} \, \bigl[ \bar{B} \, \gamma^\mu\, ( V_\mu \, , \, B) \, \bigr]
\, \, \quad {\rm and} \quad
{\rm Tr} \, \bigl[ \bar{B} \, \sigma^{\mu \nu}\, ( D_\nu V_\mu \, ,
\, B) \, \bigr] \, \, \, .
\end{equation}
Furthermore, up to order $q^4$ and for  diagrams with one baryon line
running through, because of momentum conservation we have $[D_\mu,
  V^\mu]=0$.  This argument holds e.g. for elastic forward scattering.
More generally, such terms do not contribute to the effective action
since
\begin{equation}
\partial^x_\mu \, \partial^{\lambda , y} \, G_{\nu \lambda} (x) -
\partial^x_\nu \, \partial^{\lambda , y} \, G_{\mu \lambda} (x) = 0 \, \,
\,
\end{equation}
This condition eliminates a large class of possible
terms. To arrive at the terms shown in Eq.(\ref{Piv}), one makes use
of the antisymmetry of $V_{\mu \nu}$ and uses the equations of motion
to let the covariant derivative act on the baryon fields as much as
possible. Of course, the discrete symmetries, hermiticity and so on
have to be obeyed. These are more standard in the vector field
formulation and we thus refrain from discsussing these in detail here.

The last point we want to address are the terms in ${\cal L}_{B\phi V}$
which contain the symmetric tensor $\bar{V}_{\mu \nu}$. One example is
the last term in Eq.(\ref{Piv}), another such term is e.g. given by
\begin{equation}
{\rm Tr} \, \biggl[ \bar{B} \, \gamma^\mu \, \biggl( [ D_\lambda ,
\bar{V}_{\mu \nu} ] \, , \, [D^\lambda , [D^\nu , B] ] \, \biggl)
\biggl] \, \, .
\end{equation}
In the effective action, it leads to  the following
four--point function (to be precise, the derivative of it),
\begin{eqnarray}
& &- i \,\Delta^\dagger_{\mu \nu , \rho \sigma} (x,y) = \nonumber \\
& &\partial^x_\nu \, \partial^y_\sigma \, G_{\mu \rho}(x,y) +
\partial^x_\nu \, \partial^y_\rho \, G_{\mu \sigma}(x,y) -
\partial^x_\mu \, \partial^y_\sigma \, G_{\nu \rho}(x,y) -
\partial^x_\mu \, \partial^y_\rho \, G_{\nu \sigma}(x,y)  \nonumber \\
& =&-i \int \frac{d^4 k}{(2\pi)^4} \frac{{\rm e}^{-i k \cdot (x-y)}}{k^2
- M_V^2 + i\epsilon} \, \biggl\{ g_{\mu \rho}  k_\nu k_\sigma +
g_{\mu \sigma}  k_\nu k_\rho - g_{\nu \rho}  k_\mu k_\sigma -
g_{\nu \sigma}  k_\mu k_\rho \, \biggr\} \, \, ,\nonumber \\
& & \,
\end{eqnarray}
which is symmetric under $\rho \leftrightarrow \sigma$. Therefore,
such a term or the last one in Eq.(\ref{Piv}) do not seem to have analoga
in the antisymmetric tensor field formulation. If that were to be
true, then already at tree level one would have an inequivalence
between the two representations for the massive spin--1 fields starting
at order ${\cal O}(q^4)$. However, setting (see also appendix B)
\begin{equation}
V_\nu = - \frac{1}{M_V^2} \, D^\lambda \, V_{\lambda \nu}
\end{equation}
one can rewrite $\bar{V}_{\mu \nu}$ in the following way
\begin{equation}
\bar{V}_{\mu \nu} = - \frac{1}{M_V^2} \, \bigl( D_\mu \, D^\lambda \,
V_{\lambda \nu} +  D_\nu \, D^\lambda \, V_{\lambda \mu} \, \bigr)
\,\, \, .
\end{equation}
This latter object is up to an irrelevant prefactor similar to the
tensor field combination
\begin{equation}
 D_\mu \, D^\lambda \, W_{\lambda \nu} +
D_\nu \, D^\lambda \, W_{\lambda \mu} \, \,\, \, ,
\end{equation}
which after using the equations of motion for $W_{\mu \nu}$ can be
recast into the form
\begin{equation}
 2 D_\mu \, D^\lambda \, W_{\lambda \nu} +  M_V^2 \, W_{\mu \nu} \,
\,\, \, .
\label{equiapp}
\end{equation}
Consequently, if we decompose $\bar{V}_{\mu \nu}$ as
\begin{equation}
 \bar{V}_{\mu \nu} = 2 \, D_\mu \, V_\nu  -  V_{\mu \nu} \,
\,\, \, ,
\label{a16}
\end{equation}
one realizes that the seemingly inequivalent terms containing the
symmetric tensor $\bar{V}_{\mu \nu}$ are also appearing in the tensor
field formulation for  appropriate combinations of $W_{\mu \nu}$ and
the covariant derivatives (like e.g.
given in Eq.(\ref{equiapp})).

\vfill


\setcounter{equation}{0}
\section{On the Equivalence  from Dual Transformations}
\label{appb}

Here, we wish to critically examine the approach of Ref.\cite{bp}, were it was
argued that the equivalence between the  tensor and vector field formulations
can be shown formally by using path integral methods and a set of dual
transformations as motivated by e.g. Refs.\cite{dual1},\cite{dual2}.
The starting point of Ref.\cite{bp}
is the generating functional in the presence of Goldstone bosons and
external vector and axial--vector sources,
\begin{equation}
{\cal Z}[u_\mu , f^+_{\mu \nu}] = \int [DV_\mu] \, \delta(\partial_\mu V^\mu)
{\rm e}^{i \int d^4x \, {\cal L}_V } \, \, \, ,
\label{bpstart}
\end{equation}
where ${\cal L}_V$ denotes the meson Lagrangian of Goldstone bosons chirally
coupled to massive spin--1 fields in the vector formulation. First, we wish to
examine the role of the $\delta$--function which ensures transversality of the
vector field. From the free part of the Lagrangian, Eq.(\ref{lvfree}),
we derive the
equation of motion (eom), the well--known Proca equation,
\begin{equation}
\Box \, V_\nu^a - \partial_\nu \, (\partial \cdot V^a) + M_V^2 \, V_\nu^a = 0
\, \, \, ,
\label{proca}
\end{equation}
which for $M_V \neq 0$ clearly leads to the transversality condition, $\partial
\cdot V^a = 0$.  Let us now examine the generating functional more closely.
Denote by  $V_\mu^0$ the field which fulfills the equation of motion
(we drop the flavor index $'a'$). One finds for the effective action
expanded aroung the classical solution
\begin{eqnarray}
S[V_\mu] &=& \int d^4x \, {\cal L}_V= S[V_\mu^0] + \int d^4 x \,
(V_\mu - V_\mu^0) \, \frac{\delta S}{\delta V_\mu}\biggl|_{V_\mu^0}
\nonumber \\
& + & \frac{1}{2} \,  \int d^4 x \, d^4 y \, (V_\mu - V_\mu^0) (x) \,
(V_\nu - V_\nu^0) (y) \,
\frac{\delta^2 S}{\delta V_\mu \delta V_\nu} (x,y)\biggr|_{V_\mu^0} \, \, ,
\end{eqnarray}
which using the eom leads to the generating functional
\begin{equation}
{\cal Z}[u_\mu, f^+_{\mu \nu}] = \exp\bigl(i S[V_\mu^0] \bigr) \, {\rm det}
({\cal M})^{-1/2} \, \, \, ,
\end{equation}
with ${\cal M} = (-i/2)[(\Box+M_V^2)g^{\mu \nu} - \partial^\mu \partial^\nu]$.
Neglecting loops with spin--1 mesons, we have det$\, {\cal M} =1$ and arrive at
\begin{equation}
{\cal Z}[u_\mu, f^+_{\mu \nu}] = \exp\bigl(i S[V_\mu^0] \bigr)
= \int [DV_\mu] \, \delta(V_\mu - V_\mu^0) \,
{\rm e}^{i \int d^4x \, {\cal L}_V} \, \, \, ,
\label{bmstart}
\end{equation}
which differs from Eq.(\ref{bpstart}) by the argument of the
$\delta$--function. To further sharpen this constrast, let us perform the
following substitution
\begin{equation}
V^\mu \to {V^\mu}' = V^\mu + \frac{f_V}{\sqrt{2}M_V^2} \partial_\nu f_+^{\nu
\mu} + \frac{i g_V}{\sqrt{2}M_V^2} \partial_\nu [u^\nu , u^\mu ] \, \, \, ,
\label{trafo}
\end{equation}
which also defines a transverse field and differs in the eom from the original
one by terms of order ${\cal O}(q^5)$ (and higher). To the order we are working
and for the processes we are considering, one has
\begin{equation}
{\cal L} (V_\mu ) = {\cal L} ( {V_\mu}' )
\end{equation}
so that the generating functional can also be cast in the form
\begin{equation}
{\cal Z}[u_\mu, f^+_{\mu \nu}]
= \int [DV_\mu] \, \delta\bigl( (\Box+M_V^2) \, V_\mu
- \partial_\mu \partial \cdot V \bigr)
\, {\rm e}^{i \int d^4x \, {\cal L}_V}
\, \, \, .
\label{bmtrafo}
\end{equation}
The two representations, Eq.(\ref{bmstart}) and Eq.(\ref{bmtrafo}), are
obviously equivalent as long as $V_\mu$
and $V_\mu '$ fulfill the  respective eom. This also ensures
 tranversality. However, the argument can not be reversed, not any
field which  obeys $\partial \cdot V =0$ is a solution of the eom. This is in
contrast to the starting point of Ref.\cite{bp}, Eq.(\ref{bpstart}).

We can also make contact to the tensor field formulation. We observe that
\begin{eqnarray}
\delta\bigl( (\Box+M_V^2) \, V_\mu - \partial_\mu \partial \cdot V \bigr)
&=& \delta ( \partial^\nu V_{\nu \mu} + M_V^2 \, V_\mu ) \nonumber \\
& \sim &
\delta \biggl( V_\mu + \frac{1}{M_V^2} \, \partial^\lambda \, V_{\lambda \mu}
\biggr) \, \, \, .
\label{delta}
\end{eqnarray}
Consider now $V_{\mu \nu}$ as the fundamental degree of freedom (instead of
$V_\mu$). This leads to the following generating functional\footnote{The sign
in the exponential is derived from $Z[u_\mu,f^+_{\mu \nu}]^*  =
Z[u_\mu,f^+_{\mu \nu}]$ which can always be achieved by an appropriate choice
of the phase of the in--state $|0\, , \, {\rm in}>$.}
\begin{equation}
\exp \, \bigl( -i \, S[V^0_{\mu \nu}] \, \bigr) =
\int [DV_\mu] \, \delta\biggl( V_\mu + \frac{1}{M_V^2} \, \partial^\lambda \,
V_{\lambda \mu}\biggr) \, {\rm e}^{-i \int d^4x \, {\cal L}_V} \, \, ,
\end{equation}
with (it is straightforward to generalize this argument and include
the covariant instead of the partial derivative, $\partial_\mu \to D_\mu$)
\begin{eqnarray}
{\cal L}[V_{\mu \nu}] = -\frac{1}{4} {\rm Tr} \,
\biggl( V_{\mu \nu} V^{\mu \nu}
\biggr) &+& \frac{1}{2M_V^2} \, {\rm Tr} \, \biggl(
D_\lambda V^{\lambda \mu}   D^\rho V_{\rho \mu} \biggr)
- {\rm Tr} \, \biggl( V_{\mu \nu} \, J^{\mu \nu}
\biggr) \,  . \nonumber \\
\end{eqnarray}
We make the field redefinition
\begin{equation}
\tilde{W}_{\mu \nu} \equiv \frac{1}{M_V} \, V_{\mu \nu}
\label{redef}
\end{equation}
where $\tilde{W}_{\mu \nu} $ fulfills the following eom,
\begin{equation}
\partial^\mu \, \biggl\{ \partial_\mu \partial^\rho \, \tilde{W}_{\rho \nu}
- \partial_\nu \partial^\rho \, \tilde{W}_{\rho \mu}  +M_V^2 \,
\tilde{W}_{\mu \nu} \, \biggr\} = 0 \, \, \, .
\label{eomwtil}
\end{equation}
The term in the curly brackets is nothing but the eom for the
antisymmetric tensor field, cf. the appendix of Ref.\cite{reso}. This
can be used to show the formal (symbolic) equivalence
\begin{equation}
{\cal Z}[u_\mu, f^+_{\mu \nu}]
= \exp \, \bigl(\, i \, S[V_\mu^0] \, \bigr)
= \exp \, \bigl(\, i \, S'[\tilde{W}_{\mu \nu}^0] \, \bigr) \, \, \, ,
\label{formeq2}
\end{equation}
provided one substitutes the vector field and its respective field
strength tensor in the Lagrangian as follows
\begin{equation}
{\cal L}_V [V_\mu , V_{\mu \nu}] \to
-{\cal L}_V \biggl[-\frac{1}{M_V} D^\lambda \tilde{W}_{\lambda \mu}
, M_V \, \tilde W_{\mu \nu} \biggr] \, \, \, .
\label{lequiv}
\end{equation}
Therefore, the rather conventional treatment of the path integral presented
here leads to
\begin{equation}
{\cal Z}[u_\mu, f^+_{\mu \nu}]
= \int [DV_\mu] \, \delta\biggl(  \, V_\mu +\frac{1}{M_V^2} \, D^\lambda
V_{\lambda \mu} \, \biggr) \, {\rm e}^{\, i \int d^4x \, {\cal L}_V}
\, \, \, .
\label{bmfinal}
\end{equation}
In Ref.\cite{bp}, an extra integration over the field $H_{\mu \nu}$,
$\int [DH_{\mu \nu}]$, is performed.  This additional
integration allows to produce additional constant terms in the Goldstone
boson--spin--1 Lagrangian and is at the heart of the formal
equivalence. However, the coefficients of the local contact terms can
only be fixed since it is known that the tensor formulation indeed
gives the correct result (for e.g. the pion radius). We
remark here that in the papers on dual transformations, the components
of the field strength tensor are in general not treated as a degrees
of freedom independent of the vector field components.
Let us summarize our observations. Starting from
Eq.(\ref{bmfinal}) and considering $V_{\mu \nu}$ as the fundamental object, we
could introduce (by rescaling) the field ${\tilde W}_{\mu \nu}$ which
obeys the eom of
the antisymmetric tensor field. Furthermore, the Lagrangian can be shown to be
identical to the one in tensor field formulation. This allows to identify
${\tilde W}_{\mu \nu}$  with the fundamental antisymmetric tensor field
$W_{\mu \nu}$ up to an overall constant which can be absorbed in the coupling
constants appearing in
the current $J_{\mu \nu}$. This equivalence is, however, formal.
The complete equivalence as discussed in Ref.\cite{bp} can only be achieved if
one treats the vector field $V_\mu$ and the vector $D^\rho H_{\rho \mu}$
appearing in the dual transfomation as independent degrees of
freedom. Again, this does not fix the coefficients of the local
contact terms, for doing that, arguments like in Ref.\cite{reso1}
have to be invoked.

\vspace{1cm}

\end{document}